\documentclass[%
preprint,%
 amssymb, amsmath,%
pre
]{revtex4-1}

\usepackage[pdftex]{graphicx,color}
\usepackage{bm}%
\usepackage{ulem}%
\usepackage{here}%

\newcommand{\karman}{K\'arm\'an}

\expandafter\ifx\csname package@font\endcsname\relax\else
 \expandafter\expandafter
 \expandafter\usepackage
 \expandafter\expandafter
 \expandafter{\csname package@font\endcsname}%
\fi
\begin{document}

\preprint{aaa/bbb}

\title{Jacobian-free algorithm to calculate the phase sensitivity function in the phase reduction theory and its applications to K\'arm\'an's vortex street}

\author{Makoto Iima}%
	\email{iima@hiroshima-u.ac.jp}
	\affiliation{Graduate School of Science, Hiroshima University, 1-7-1, Kagamiyama Higashi-Hiroshima, Hiroshima 739-8251, Japan}
\date{\today}%

\begin{abstract}
Phase reduction theory has been applied to many systems with limit cycles; however, it has limited applications in incompressible fluid systems. This is because the calculation of the phase sensitivity function, one of the fundamental functions in phase reduction theory, has a high computational cost for systems with a large degree of freedom. Furthermore, incompressible fluid systems have an implicit expression of the Jacobian. To address these issues, we propose a new algorithm to numerically calculate the phase sensitivity function. This algorithm does not require the explicit form of the Jacobian along the limit cycle, and the computational time is significantly reduced, compared with known methods. Along with the description of the method and characteristics, two applications of the method are demonstrated. One application is the traveling pulse in the FitzHugh Nagumo equation in a periodic domain and the other is the K\'arm\'an's vortex street. The response to the perturbation added to the K\'arm\'an's vortex street is discussed in terms of both phase reduction theory and fluid mechanics.
\end{abstract}

\pacs{aaa}

\maketitle
\section{Introduction}\label{sec:introduction}
Periodic flow is the simplest type of unsteady flow. A well-known example is the K\'arm\'an's vortex street observed in the downstream of a cylinder in a uniform flow, when $Re \gtrsim 50$, where $Re = UD/\nu$; $U$ is the velocity of the uniform flow, $D$ is the diameter of the cylinder, and $\nu$ is the kinematic viscosity\cite{williamson96_vortex_dynam_cylin_wake}. If the whole system is described in the phase space, the periodic flow is represented by the stable limit cycle (LC), a periodic orbit with a unique period such that all the orbits near the LC converge to LC\cite{Strogatz1994}.

The state near the LC can be described by a single variable called ``phase'', $\phi$, and the dynamics near the LC can be described by the ordinary differential equation (ODE) of $\phi$, which is a significant reduction of the degree of the freedom\cite{%
Kuramoto1984%
}
. The phase reduction technique has been applied to many problems, including mechanical vibrations, spiking neurons, and flashing fireflies\cite{nakao15_phase_reduc_approac_to_synch_nonlin_oscil}. In the phase reduction theory, the phase sensitivity function, which gives the linear response coefficients of the phase to perturbations, is essential for the reduction of the original system into the ODE system of $\phi$
\cite{Kuramoto1984,%
nakao15_phase_reduc_approac_to_synch_nonlin_oscil}. A practical method to calculate the phase sensitivity function, ``the adjoint method''\cite{ermentrout96_type_i_membr_phase_reset_curves_synch}, has been applied to many problems not only in the systems described by the ordinary differential equations\cite{nakao15_phase_reduc_approac_to_synch_nonlin_oscil}, but also the partial differential equations, including the reaction-diffusion system\cite{nakao14_phase_reduc_approac_to_synch}, convection in Hele-Shaw cell\cite{kawamura15_phase_descr_oscil_convec_with}, and flagellum synchronization of a model equation\cite{kawamura18_phase_reduc_approac_to_elast}.

If the phase reduction theory is applied to incompressible fluid systems, not only are new aspects of the periodic flows obtained, but also new techniques of the flow control will be developed. In the case of the K\'arm\'an's vortex street, we will be able to design an efficient perturbation form to change the phase of the flow (or the timing of the separation of vortices). Further, the synchronization of two cylinders
\cite{%
williamson85_evolut_singl_wake_behin_pair_bluff_bodies,%
peschard96_coupl_wakes_cylin,%
akinaga05_linear_stabil_flow_past_two%
} can be analyzed.

However, the phase reduction theory has not been applied to incompressible fluids, except for the cases in which the linearized equation around the LC can be obtained explicitly\cite{kawamura15_phase_descr_oscil_convec_with} and the case of direct measurements of the phase shift by adding perturbation to the flow to obtain the information of the prescribed regions\cite{taira18_phase_respon_analy_synch_period_flows}. A practical problem exists, if the phase reduction theory is applied to address incompressible fluids, which is, a computational source. The description of the states near the LC requires the linearized matrix (the Jacobian) of any point along the LC; further, the explicit form of the matrix cannot be obtained because the Poisson equation must be solved to obtain the pressure. In such a case, we need to numerically calculate all the elements of the Jacobian and store the values, which requires a lot of memory. Recently, Taira and Nakao\cite{taira18_phase_respon_analy_synch_period_flows} calculated the phase sensitivity function at the top of the separation point of the cylinder in a uniform flow by a direct method. However, a computational source is needed to obtain the phase sensitivity function of the whole region. Further, it requires a lot of periods for the perturbed system to converge to LC. They will be discussed later.

In this paper, we propose a new numerical algorithm to calculate the phase sensitivity function, which is applicable for the systems with large degree of freedom and without explicit expression of the Jacobian. In particular, we applied this method to analyze the K\'arm\'an's vortex street. The obtained phase sensitivity function revealed that the downstream region comprises narrow bands of the effective area for the phase shift, and the distribution and effective directions are time-dependent. Moreover, the distribution in the upstream region is less time-dependent, which suggests that controlling the phase is more convenient. Furthermore, a comprehensive interpretation based on fluid mechanics is discussed.

The remainder of this paper is organized as follows. In Sec. \ref{sec:Methods}, we describe the details of the proposed method and discuss the characteristics. In Sec. \ref{sec:Applications}, we demonstrate the proposed method. In Sec. \ref{sec:A traveling pulse in Fitz-Hugh Nagumo equation in a periodic domain}, the phase sensitivity function of the traveling pulse is compared with known results. In Sec. \ref{sec:Karman's vortex street}, we exhibit the phase sensitivity function of the K\'arm\'an's vortex street and discuss the detailed characteristics in terms of fluid mechanics. In Sec. \ref{sec:Summary}, we summarize the results.

\section{Methods}
\label{sec:Methods}
In this section, the proposed method is described in detail to numerically calculate the phase sensitivity function of a dynamic system with finite dimensions. In the case of the partial differential equations (PDE), the discretized system is considered. 
\subsection{Phase reduction}
\label{sec:Phase reduction}
The definitions and notations of the phase reduction theory are briefly summarized in this section. Refs.\cite{nakao15_phase_reduc_approac_to_synch_nonlin_oscil,%
Kuramoto1984%
} contains more information on this theory. Let us consider the $n$-dimensional autonomous dynamical systems given by:
\begin{eqnarray}
 \frac{d\bm{x}}{dt} = \bm{f}(\bm{x}),
 \label{eq:dynamical system}
\end{eqnarray}
 where the vector $\bm{x}={}^t(x_1,\cdots,x_n) \in \mathbb{R}^n$ represents the state, and the function $\bm{f}(\bm{x})={}^t(f_1(\bm{x}),\cdots,f_n(\bm{x})) : \mathbb{R}^n \mapsto \mathbb{R}^n$ determines the dynamics. It is assumed that eq. (\ref{eq:dynamical system}) has a stable limit cycle solution $\bm{x}(t)=\bm{p}(t)$ that satisfies $\bm{p}(t+T)=\bm{p}(t)$ for all  $t$, where $T$ is the natural period.

In the phase reduction theory, a value called ``phase'', $\phi$, (more precisely, the ``asymptotic phase'' in Ref.\cite{nakao15_phase_reduc_approac_to_synch_nonlin_oscil}) is defined on and near the limit cycle (LC) as follows: on the LC, the origin of the phase ($\phi=0$) is chosen, and the orbit $\bm{p}(t)$ that starts from the origin at $t=0$ is considered; then, $\phi$ is defined as $\displaystyle \phi(t) = \omega t\; (\mod 2\pi)$, where $\omega= 2\pi/T$ and $0 \le \phi(t) < 2\pi$. For the phase at a particular point $\bm{x}_{0}$ near the LC, $\Phi(\bm{x}_0)$, the orbit $\bm{x}(t)$ that starts from $\bm{x}=\bm{x}_0$ at $t=0$ is considered. Now, $\Phi(\bm{x}_0)=\phi_0$, in which, $\phi_0$ satisfies $\lim_{t \to \infty} (\bm{x}(t)-\bm{p}(t+\phi_0/\omega))= 0$.

A fundamental function of the phase reduction theory is the phase sensitivity function,  $\bm{Z}(\phi)$, which is defined as:
\begin{equation}
\bm{Z}(\phi) = 
\left.
\frac{\partial \Phi(\bm{x})}{\partial \bm{x}}
\right|_{\Phi(\bm{x})=\phi}.
\end{equation}
The phase sensitivity function determines the phase increment of the state near the LC due to a small perturbation $\Delta \bm{x}$, because of the relationship $\displaystyle \Phi(\bm{x}+\Delta \bm{x}) =\bm{Z}(\phi) \cdot \Delta \bm{x} + O(|\Delta \bm{x}|^2)$. Once $\bm{Z}(\phi)$ is obtained, the extent of phase shift after perturbation due to any force, and the synchronized property of the coupled oscillators, can be calculated\cite{%
winfree67_biolog_rhyth_behav_popul_coupl_oscil,%
Kuramoto1984,%
nakao15_phase_reduc_approac_to_synch_nonlin_oscil%
}.

If $\displaystyle \bm{\tilde{Z}}(t) = (1/\omega)\bm{Z}(\omega t)$ is defined, $\bm{Z}(\phi)$ is obtained as the periodic solution of the adjoint equation as follows: 
\begin{equation}
  \frac{d \bm{\tilde{Z}}}{dt} = - {}^tJ(\bm{p}(t))\bm{\tilde{Z}},  
  \label{eq:adjoint equation}
\end{equation}
 where $J$ is the Jacobian of the dynamical system (\ref{eq:dynamical system}), and its $(i,j)-$component is $J_{ij} = \frac{\partial f_i}{\partial x_j}(\bm{p}(t))$
\cite{ermentrout96_type_i_membr_phase_reset_curves_synch}. Because the adjoint equation is linear, an additional condition to determine the amplitude of $\tilde{\bm{Z}}(t)$ is required. The following normalization relationship is imposed:
\begin{equation}
 \tilde{\bm{Z}}(t)\cdot\bm{f}(\bm{p}(t)) = 1.
\label{eq:normalization of Z}
\end{equation}
Once $\tilde{\bm{Z}}(t)$ is obtained, $\bm{Z}(\phi)$ can be easily obtained by converting $t$ to $\phi$.

\subsection{Problems in calculations of the phase sensitivity function in the incompressible fluid system}
\label{sec:Problems in calculations of the phase sensitivity function in incompressible fluid system}
Let us consider the case that the dimension $n$ is large and the function $\bm{f}(\bm{x})$ is not given explicitly, which is the case of the discretized system of the incompressible fluid system for numerical calculation.

A simple way to calculate the phase shift due to the perturbation is called ``direct method''\cite{nakao15_phase_reduc_approac_to_synch_nonlin_oscil,%
taira18_phase_respon_analy_synch_period_flows%
}. A phase shift $\Delta\Phi$ due to the perturbation $\Delta \bm{x}$ added to the state $\bm{x}$ is given by:
\begin{equation}
  \Delta \Phi(\bm{x}, \Delta \bm{x}) = \Phi(\bm{x}+\Delta \bm{x})-\Phi(\bm{x}).
\end{equation}
Then, the phase sensitivity function $\bm{Z}(\phi)=(Z_1(\phi),\cdots,Z_n(\phi))$ is obtained by:
\begin{equation}
  Z_j(\phi) = \lim_{\epsilon \to 0} \frac{\Delta \Phi(\bm{x}, \epsilon \bm{e}_j)}{\epsilon},
\end{equation}
 where $\bm{e}_j$ is a unit vector and its $k-$th component is $[\bm{e}_j]_k = \delta_{jk}$ ($\delta_{jk}$ is the Kronecker's delta).

To evaluate the calculation time by the direct method, $\Delta \tau$ is assumed as the time to calculate one time step of the dynamical system (\ref{eq:dynamical system}) per one degree of freedom, i.e., time $n\Delta \tau$ is needed to calculate single time step of the whole system. It is further assumed that the time step is given by the $T/m$, where $m$ is the division number of the period $T$ and that it needs $N_1$ periods for the system to converge to estimate the phase shift $\Delta \Phi$. Then, $N_1m$ steps (equivalently, $N_1$ periods) are needed to calculate the phase shift due to the single perturbation $\Delta \bm{x}=\epsilon \bm{e}_j$. Therefore, the total calculation time is estimated as $N_1 mn^2\Delta \tau$ because $n$ perturbations $\left\{ \epsilon \bm{e}_j \mid j=1,\cdots,n \right\}$ are needed to obtain $\bm{Z}(\phi)$.

If the adjoint method is applied, an asymptotic state of the adjoint equation (\ref{eq:adjoint equation}) must be obtained as $t \to -\infty$. When applying this procedure to the incompressible fluid system, several problems occur. In this case, the analytic expression of the Jacobian $J$ cannot be obtained because the pressure must be calculated by solving the Poisson equation. Thus, we need to calculate $J$, $n \times n$ matrix, every time step on $\bm{x}(t)=\bm{p}(t)$. The memory required to store the Jacobians along the LC is estimated of the order of $mn^2$. For example, approximately 800 GB is required to store $J(\bm{p}(t))$ with double precision, when $m=1,000$ and $n=10,000$, which is too large for the calculation.

Alternatively, the components of $J(\bm{p}(t))$ can be evaluated at each time step. If the following formula is used,
\begin{eqnarray}
  J_{ij}(\bm{p}(t)) = \frac{\partial f_i}{\partial x_j}(\bm{p}(t))
 = \lim_{\epsilon \to 0} \frac{x_i(\bm{p}(t)+\epsilon \bm{e}_j)-x_i(\bm{p}(t))}{\epsilon},
\label{eq:appximation of L}
\end{eqnarray}
 the calculation time of $J$ at a particular time is of the order $n^2\Delta \tau$. If the adjoint system takes $N_2$ periods to converge, we need an order of time $N_2 mn^2 \Delta \tau$, which is the same order of the calculation time for the direct method if we can assume that $N_1$ and $N_2$ are of the same order.
These methods require long-time integration of the original system or the adjoint system until convergence. This process can be reduced by the proposed method, which is discussed in the next subsection.

\subsection{Proposed method}
\label{sec:Proposed method}
We propose a novel method to calculate the phase sensitivity function that can reduce the computational cost by the factor $1/N_1$ or $1/N_2$.

In this method, a Jacobian-free method is utilized to calculate the product $J \bm{v}$,
 where $\bm{v}$ is a vector, by the following formula:
\begin{equation}
  J(\bm{p}(t)) \bm{v} = \lim_{\epsilon \to 0} \frac{\bm{x}(\bm{p}(t)+\epsilon \bm{v})-\bm{x}(\bm{p}(t))}{\epsilon}
\label{eq:jacobian-free method for Lv}
\end{equation}
\cite{knoll04_jacob_free_newton_krylov_method}.
This method can minimize the time required because the $n$ components of $J\bm{v}$ can be calculated at one time. If we need to calculate all the components of $J$ by eq. (\ref{eq:appximation of L}), the computational time of $J \bm{v}$ is of the order of $n^2\Delta \tau$, while it is $n\Delta \tau$ when eq. (\ref{eq:jacobian-free method for Lv}) is used. The formula (\ref{eq:jacobian-free method for Lv}) cannot be applied for the calculation of the adjoint equation (\ref{eq:adjoint equation}) because no similar formula has been known for ${}^tJ \bm{v}$
\cite{Govaerts2000}.

Now, we consider a method to obtain the periodic solution of the adjoint equation (\ref{eq:adjoint equation}). The linearized equation of (\ref{eq:dynamical system}) from the limit cycle $\bm{p}(t)$ can be written as: 
\begin{equation}
  \frac{d\bm{y}}{dt} = A(t) \bm{y},\quad
  A(t) = J(\bm{p}(t)).
  \label{eq:linearized equation}
\end{equation}
The linearly independent set of the solution of (\ref{eq:linearized equation}), $\bm{y}_1, \cdots \bm{y}_n$, can be used to define the fundamental solution matrix as follows:
\begin{equation}
G_p(t) = \left(
\bm{y}_1(t) \; \cdots \; \bm{y}_n(t)
\right).
\end{equation}
Then, the general solution $\bm{y}(t)$ is represented by $\bm{y}(t) = G_p(t) \bm{c}$, where $\bm{c}$ is a constant column vector determined by the initial condition.

Differentiating the identity $G_p(t)^{-1}G_p(t)=I$ with respect to $t$ and using the relationship $\displaystyle \frac{dG_p}{dt}=AG_p$, we obtain the following equation:
\begin{equation}
  \frac{d \hat{G}_p(t)}{dt} =- {}^t A \hat{G}_p(t),
\label{eq:adjoint equation 2}
\end{equation}

 where $\hat{G}_p(t) = {}^t(G_p(t)^{-1})$. Equation (\ref{eq:adjoint equation 2}) shows that $\hat{G}_p(t)$ is the fundamental solution matrix of the adjoint equation (\ref{eq:adjoint equation}). Let us assume that Eq. (\ref{eq:adjoint equation}) has a unique limit cycle solution $\bm{z}_p(t)$. Then, the periodicity condition, $\bm{z}_p(t+T)=\bm{z}_p(t)$ can be reduced to:
\begin{eqnarray}
{}^t\bm{z}_p(t) (G_p(t+T)-G_p(t)) = 0.
\label{eq:periodic solution of adjoint equation}
\end{eqnarray}
Eq. (\ref{eq:periodic solution of adjoint equation}) can be shown as follows: If we write $\bm{z}_p(t)=\hat{G}_p(t)\bm{c}$, the periodicity condition is, $\hat{G}_p(t+T) \bm{c} = \hat{G}_p(t)\bm{c}$. Taking the transpose of this equation and using the identity $G_p(t)^{-1}G_p(t)=I$, either ${}^t\bm{c} G_p(t)^{-1} G_p(t) G_p(t+T)^{-1} =  \bm{c} {}^t G_p(t)^{-1}$, or the following equation can be obtained:
\begin{equation}
 {}^t\bm{z}_p(t) (G_p(t)G_p(t+T)^{-1}-I) = 0.
\end{equation}
This equation is equivalent to Eq. (\ref{eq:periodic solution of adjoint equation}). Solving eq. (\ref{eq:periodic solution of adjoint equation}), we can obtain $\bm{z}_p(t)$, which is proportional to $\tilde{\bm{Z}}(t)$.

The calculation procedure of $\bm{z}_p(t)$ using Eq. (\ref{eq:periodic solution of adjoint equation}) is as follows. Let us rewrite $G_p(t+T)-G_p(t)$, using column vectors $\bm{g}_k(t)\;(k=1,\cdots,n)$ as follows:
\begin{equation}
  G_p(t+T)-G_p(t)= \left( \bm{g}_1(t), \bm{g}_2(t), \cdots \bm{g}_n(t) \right).
\label{eq:Floquet matrix}
\end{equation}
Then, the equation (\ref{eq:periodic solution of adjoint equation}) is decomposed to $n$ orthogonality relationships between $\bm{z}_p$ and $\{\bm{g}_k \mid k=1,\cdots,n \}$, i.e.,
\begin{eqnarray}
  {}^t\bm{z}_p(t) \bm{g}_k(t)=0.\quad (k=1,\cdots n).  
\label{eq:orthogonality equation}
\end{eqnarray}

Further, the vectors $\{\bm{g}_k\}$ can be obtained through time integration of the original system. Let us write $\bm{x}(t;t_0, \bm{x}_0)$ as the solution of Eq. (\ref{eq:dynamical system}) with $\bm{x}(t_0)=\bm{x}_0$. Then, for any small vector $\bm{y}_0$, we obtain the following relationship, if higher order terms are negligible:
\begin{eqnarray}
&&\bm{x}(t_0+T; t_0, \bm{p}(t_0)+\bm{y}_0)-\bm{x}(t_0+T; t_0, \bm{p}(t_0))-\bm{y}_0\nonumber \\
&=& \int_{t_0}^{t_0+T} J(\bm{p}(t)) \bm{y}_0 dt + O(|\bm{y}_0|^2) \nonumber\\
&\simeq& (G(t_0+T)-G(t_0)) \bm{y}_0.
\label{eq:Algorithm of Gy}
\end{eqnarray}
This formula can be used to calculate $\bm{g}_k(t_0)$ by setting $\bm{y}_0=\epsilon \bm{e}_k$, where $\epsilon$ is a small parameter. Because of the existence and uniqueness of the periodic solution, one eigenvalue of $G_p(t_0+T)-G_p(t_0)$ is zero. Therefore, $\bm{g}_1, \bm{g}_2, \cdots \bm{g}_n$ are linearly dependent, and $(n-1)$ elements of $\{ \bm{g}_k \}$ are linearly independent. Thus, there exists a non-trivial solution of (\ref{eq:periodic solution of adjoint equation}).

The algorithm to find the solution is as follows: using the Gram-Schmidt orthonormalization and applying the relationship (\ref{eq:Algorithm of Gy}) with the linearly independent set of $\bm{y}_0$, we can obtain $(n-1)$ vectors $\{\bm{a}_k \mid k=1,\cdots,n-1 \}$ which construct the orthonormal basis of the linear space $V$ spanned by $\{\bm{g}_k \mid k=1,\cdots,n \}$, i.e., $(\bm{a}_i, \bm{a}_j)=0\; (i\ne j, 1\le i,j \le n-1)$ and $|\bm{a}_i|=1$. Finally, using a general vector $\bm{y} \in \mathbb{R}^n$ such that $\bm{y} \notin V$, the following $\bm{z}$ is obtained:
\begin{equation}
  \bm{z} = \bm{y} - \sum_{k=1}^{n-1} (\bm{a}_k, \bm{y})\bm{a}_k.
\end{equation}
By definition, $\bm{z}$ is perpendicular to any vector of $V$, i.e., $(\bm{z}, \bm{a}_i)=0\; (i=1,\cdots, n-1)$. Thus, $\bm{z}$ is the solution of eq. (\ref{eq:orthogonality equation}) and eq. (\ref{eq:adjoint equation}).

\subsection{Characteristics}\label{sec:characteristics}

The characteristics of the proposed method is as follows:

First, this method is memory-saving and Jacobian-free, i.e., an explicit expression of the Jacobian is not needed. This means that we do not need memory to store the Jacobian data over one period, which requires $mn^2$ variables. Second, this method is time-saving. This method enables the calculation of the phase sensitivity function by $n-1$ calculation of time evolution over one period, $T$. The total computation time is estimated as $mn(n-1) \Delta \tau \sim mn^2 \Delta \tau$, $1/N_1$ or $1/N_2$ of the direct integration of the adjoint equation. Third, this method can be efficiently implemented by parallel computation. The bottleneck of this method is the calculation of $n-1$ vectors in the form $(G_p(t+T)-G_p(t))\bm{y}_0$. Each calculation needs time integration over one period, which requires more computation time. However, each process can be independently calculated in parallel. This means that the parallel computing algorithms such as MPI method work efficiently. These three characteristics are the advantages of the proposed method.

In addition, the following remarks should be considered. First, a periodic solution data $\bm{p}(t)$ must be prepared for the calculation; there are several algorithms to obtain the periodic solution numerically, e.g., Ref.\cite{saiki07_numer_detec_unstab_period_orbit}; Second, the proposed algorithm gives $\bm{Z}(\phi)$ for single phase, which appears a disadvantage initially. When we consider the traveling wave in a periodic domain, e.g., traveling pulse in the FitzHugh Nagumo(FHN) equation\cite{nakao14_phase_reduc_approac_to_synch}, the phase sensitivity function $\bm{Z}(\phi)$ at one phase is sufficient because of the Galilean invariance of the solution. This example will be discussed in the next section. In such cases, the problem does not need to be considered. Generally, this problem can be amended, because the LHS of eq. (\ref{eq:Floquet matrix}) for different value of $t$ can be calculated by the data of $G_p(t)$ over two periods, which does not change the order of the calculation cost.
\section{Applications}\label{sec:Applications}
The proposed method is applied to two PDE problems. First, the phase sensitivity function of a traveling pulse is calculated in the FHN equation in a periodic domain to compare with the results given in Ref.\cite{nakao14_phase_reduc_approac_to_synch}. Second, the phase sensitivity function of the K\'arm\'an's vortex street is calculated.
\subsection{A traveling pulse in FitzHugh Nagumo equation in a periodic domain}\label{sec:A traveling pulse in Fitz-Hugh Nagumo equation in a periodic domain}
\begin{figure}[h]
\centering
\includegraphics[width=0.8\textwidth]{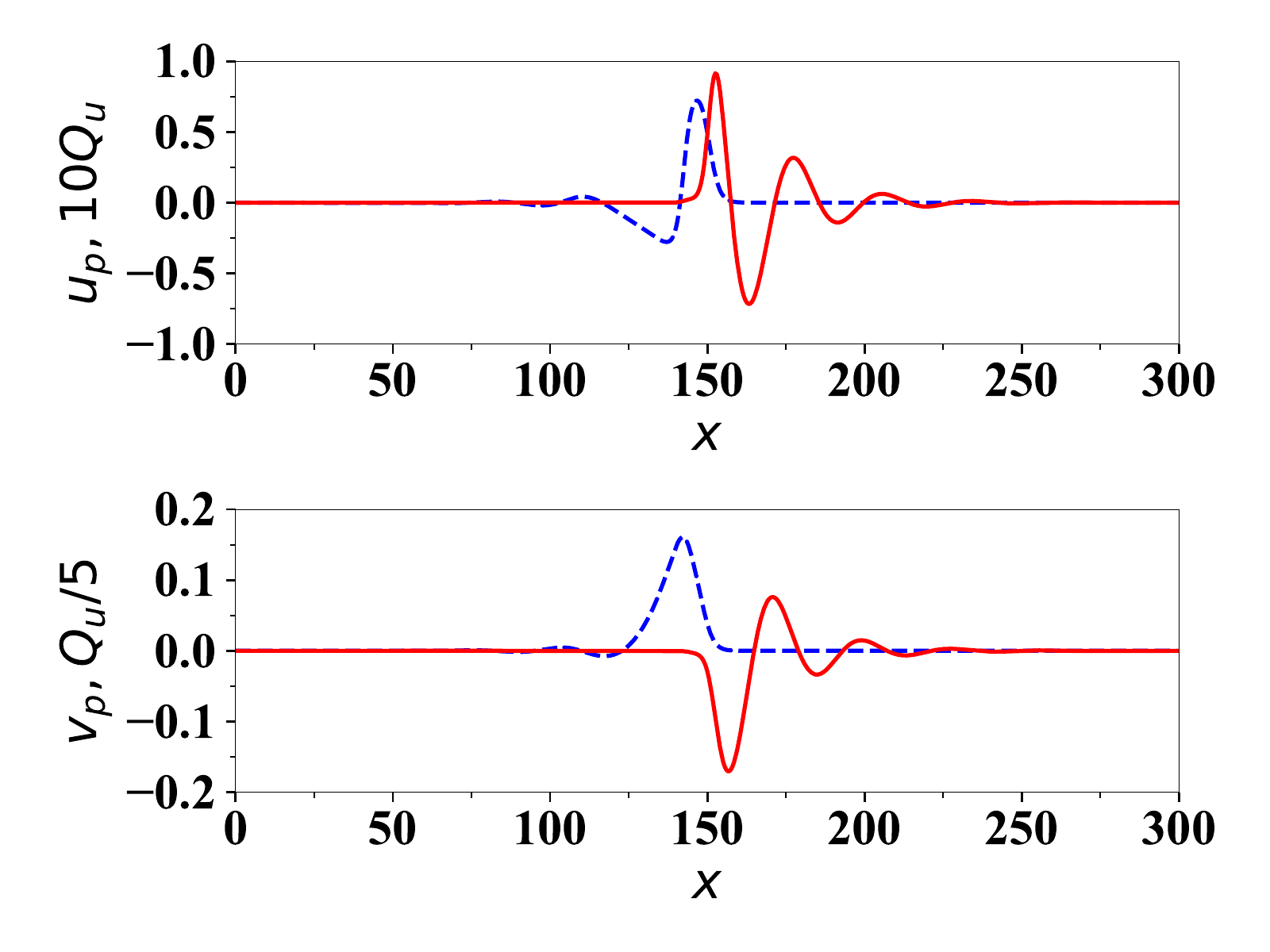}
\caption{Snapshot of the traveling pulse solution $u_p(x,\phi/\omega)$ and $v_p(x,\phi/\omega)$ with a wavy tail (blue broken lines) and the corresponding phase sensitivity functions $Q_u(x,\phi)$ and $Q_v(x,\phi)$ (red solid lines).
}
\label{fig:PSF for FHN eq}
\end{figure}

Nakao, Yanagita, and Kawamura developed phase reduction theory for the reaction-diffusion system, and calculated the phase sensitivity function for several solutions including the traveling pulse of FHN model (sec. IIIA in their paper). The FHN equations are described as:
\begin{eqnarray}
\frac{\partial u}{\partial t} &=& f_u(u,v)+\kappa \frac{\partial^2 u}{\partial x^2}, \quad
  f_u(u,v)=u(u-\alpha)(1-u)-v,\label{185158_2Dec18} \\
\frac{\partial v}{\partial t} &=& f_v(u,v)+ \delta \frac{\partial^2 v}{\partial x^2}, \quad
  f_v(u,v)=\tau^{-1}(u-\gamma v),\label{185217_2Dec18}
\end{eqnarray}
in one dimensional space. Here $u(x,t)$ and $v(x,t)$ are independent variables and $\alpha,\gamma,\tau,\kappa,$ and $\delta$
 are constants. The details are explained in Sec. IIIA in Ref.\cite{nakao14_phase_reduc_approac_to_synch}.

When the system parameters are $\alpha=0, \tau^{-1}=0.018, \gamma=1, \kappa=1,$ and $\delta=0.02$, which are the same values in Ref.\cite{nakao14_phase_reduc_approac_to_synch}, a traveling pulse solution with wavy tails exists. We calculated the FHN equation for a one-dimensional periodic domain with the size $L=300$ with spatial grids $\Delta x=0.5$; the space was discretized to $N (=L/\Delta x)$ discrete points. The time integration of the FHN equations was calculated using the Runge-Kutta method of the second order. The periodic solution was obtained by the Newton-Raphson method\cite{saiki07_numer_detec_unstab_period_orbit}, in which the GMRES(k) method with the Jacobian-free algorithm\cite{knoll04_jacob_free_newton_krylov_method} was used for solving the linear equation. The period was divided into $m(=6,000)$ discrete states. The period of the obtained solution was $T=553.15$.

The phase sensitivity function was calculated for the discretized system with $2N$ dimensions, where:
\begin{eqnarray}
\bm{U} &=& {}^t(u_1,\cdots,u_N,v_1,\cdots,v_N), \quad (u_k,v_k)=(u(k\Delta x), v(k\Delta x)), \quad (k=1,\cdots,N) \\
\bm{F}(\bm{U}) &=& {}^t(g_1^{u},\cdots,g_N^{u},g_1^{v},\cdots,g_N^{v}),\\
(g_k^{u}, g_k^{v}) &=& \left(f_u(u_k, v_k)+\kappa \frac{u_{k+1}+u_{k-1}-2u_k}{\Delta x^2},
f_v(u_k, v_k)+\delta \frac{v_{k+1}+v_{k-1}-2v_k}{\Delta x^2} \right),
\end{eqnarray}
 with the periodic boundary condition $u_{j+N}=u_{j}, v_{j+N}=v_{j}$ for any $j$. The discretized equations for Eqs. (\ref{185158_2Dec18}) and (\ref{185217_2Dec18}) are represented as $\displaystyle
\frac{d\bm{U}}{dt}=\bm{F}(\bm{U})
$. The phase sensitivity function was converted to the phase sensitivity function for the reaction diffusion system, defined by (B13) in Ref.\cite{nakao14_phase_reduc_approac_to_synch}, given by:
\begin{equation}
   (Q_u(x;\phi), Q_v(x;\phi)) =
\left. 
\left(
 \frac{\delta \Phi(u, v)}{\delta u},
 \frac{\delta \Phi(u, v)}{\delta v}
\right)
\right|_{(u,v)=(u_p(x,\frac{\phi}{\omega}), v_p(x,\frac{\phi}{\omega}))},
\end{equation}
 where $(u_p(x,t), v_p(x,t))$ is the traveling pulse solution which is time periodic. The difference between $(Q_u(x;\phi), Q_v(x;\phi))$
 and $\bm{Z}(\phi)$ is just a numerical factor when the grid distance $\Delta x$ is homogeneous. Normalization of $(Q_u(x;\phi), Q_v(x;\phi))$ is given by (B9) in Ref. \cite{nakao14_phase_reduc_approac_to_synch}, which is:
\begin{eqnarray}
\omega &=& 
\int 
\left[
Q_u(x;\phi) \left\{ f_u(u,v)+\kappa \frac{\partial^2 u}{\partial x^2} \right\}
 + Q_u(x;\phi) \left\{ f_v(u,v)+\delta \frac{\partial^2 v}{\partial x^2} \right\}
\right]
dx \\
&\simeq& 
\sum_{k=1}^{N}
\left\{
 Q_u(k\Delta x;\phi) g_k^{u}
+Q_v(k\Delta x;\phi) g_k^{v}
\right\}\Delta x.
\label{eq:normalization of Q}
\end{eqnarray}
The normalization condition for $\bm{Z}(\phi)$, eq. (\ref{eq:normalization of Z}) reads
$\sum_{k=1}^{N} \left\{ Z_k^{u}(\phi)g_k^{u} + Z_k^{v}(\phi)g_k^{v} \right\}=1$, where:
\begin{equation}
\bm{Z}(\phi) = (Z_1^{u},\cdots,Z_N^{u}, Z_1^{v},\cdots,Z_N^{v}), \quad
Z_k^{u} = \frac{\partial \phi}{\partial u_k}, \quad
Z_k^{v} = \frac{\partial \phi}{\partial v_k}\; (k=1,\cdots,N).
\label{eq:normalizatino of Z for reaction-diffusion}
\end{equation}
Eqs. (\ref{eq:normalization of Q}) and (\ref{eq:Def of Q}) produce the following relationship:
\begin{equation}
  (Q_u(k\Delta x;\phi), Q_v(k\Delta x;\phi)) =\frac{\omega}{\Delta x}(Z_k^{u}(\phi), Z_k^{v}(\phi)).
\label{eq:Def of Q}
\end{equation}

In Fig. \ref{fig:PSF for FHN eq}, the traveling pulse solution and corresponding phase sensitivity functions $Q_u$ and $Q_v$ at a phase are shown. The shapes of both the traveling pulse solution $(u_p,v_p)$ and the phase sensitivity functions are similar to Fig. 1(a) in Ref.\cite{nakao14_phase_reduc_approac_to_synch}, which validates our proposed method. In this case, the solution is a traveling one, i.e., $u(x,t)=u(x-ct,0)$ and $v(x,t)=v(x-ct,0)$, where $c$ is the speed of the traveling pulse. Let us assume that the phase sensitivity functions in Fig. \ref{fig:PSF for FHN eq} is at $\phi=0$. Then, we have the relationships $Q_u(x,\phi) = Q_u(x-c\phi/\omega, 0)$ and $Q_v(x,\phi) = Q_v(x-c\phi/\omega, 0)$. Thus, the phase sensitivity function at another phase can be obtained by the spatial translation of $Q_u$ and $Q_v$.

Here, the traveling pulse propagates to the right. Thus, the perturbation in the right area of the pulse interacts with the entire wavy tail. Such interaction causes a relatively significant phase shift. In addition, due to the wavy characteristics, both $Q_u$ and $Q_v$ oscillate spatially. Moreover, the perturbation in the left area of the pulse only interacts with a part of the wavy tail and the phase shift is less significant. In this sense, the right area is the ``upstream'' of the pulse. The wavy shapes of $Q_u$ and $Q_v$ on the right of the pulse match these observations.


\subsection{K\'arm\'an's vortex street}\label{sec:Karman's vortex street}

\subsubsection{Methods}
\label{sec:Methods(karman)}
\begin{figure}[h]
\centering
\includegraphics[width=0.7\textwidth]{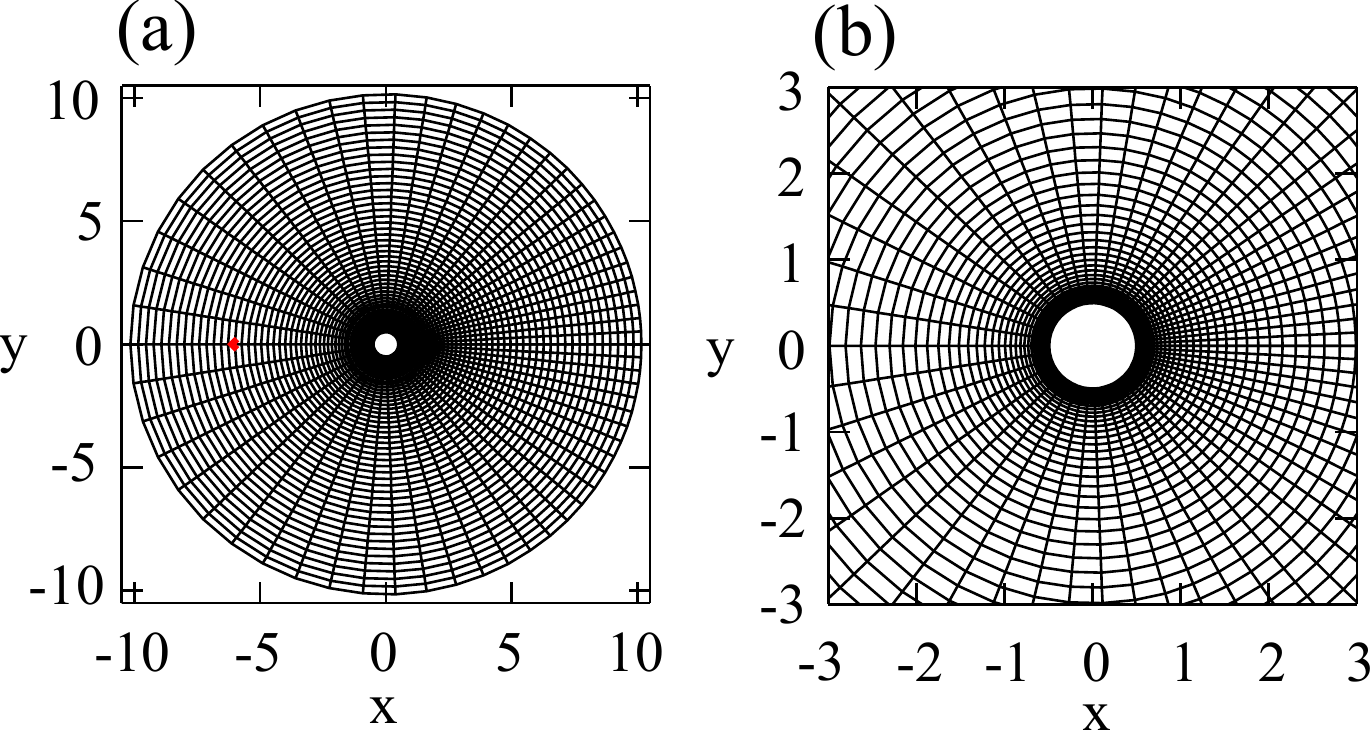}
\caption{
Computational grid for the calculation of the Karman's vortex street. (a) Whole computational region. Red point at $\bm{x}=\bm{x}_{pert}=(-6.00, 0)$ indicates the position of the perturbations to demonstrate the phase shift in Fig. \ref{fig:A demonstration of the phase shift}. (b) Magnified region near the cylinder.
}
\label{fig:Computational grid for the caculation of karman vortex street}
\end{figure}

\begin{figure}[h]
\centering
\includegraphics[width=1.1\textwidth]{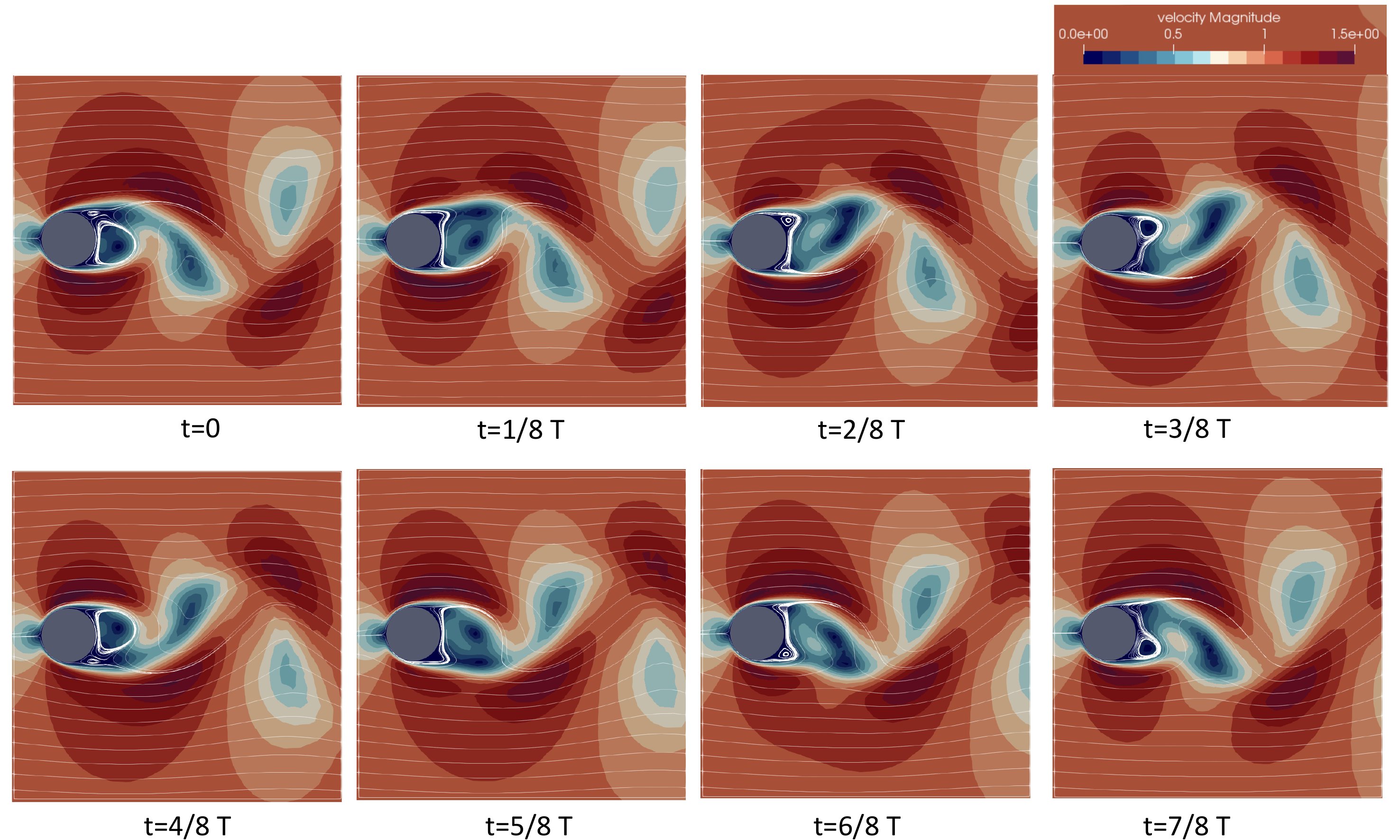}
\caption{
Snapshots of the K\'arm\'an's vortex street. Colors indicate the flow speed, and curves indicate the streamlines.}
\label{fig:Karman vortex}
\end{figure}

We consider the flow past the cylinder in a uniform flow in two-dimensional space. The flow is governed by the incompressible Navier-Stokes equations in the non-dimensional form:
\begin{eqnarray}
  \frac{\partial \bm{u}}{\partial t}
+ \bm{u}\cdot \nabla \bm{u}
= -\nabla p + \frac{1}{Re} \Delta \bm{u},
\quad
\nabla \cdot \bm{u} = \bm{0},
\label{eq:Navier-Stokes eqs}
\end{eqnarray}
where $\bm{u}=(u,v)$ is the velocity, $p$ is the pressure, and $Re$ is the Reynolds number.

The diameter of the cylinder is unity, and the uniform flow is represented by $\bm{U}=(1,0)$. The computational domain is a circle of radius $R$ and the center, which is also the center of the cylinder, is the origin of the coordinate.

The boundary conditions in the polar coordinate $(r,\theta)$ are given as follows: the non-slip condition ($\displaystyle
\bm{u}=\bm{0}$ and $\displaystyle \frac{\partial p}{\partial r}=0$) was applied to the cylinder ($\displaystyle r=\frac12$). On the outer boundary ($\displaystyle r=R$), the inflow condition ($\displaystyle
\bm{u}=(1,0)$ and $\displaystyle \frac{\partial p}{\partial r}=0$) was applied in the region $\displaystyle \frac{\pi}{3} < \theta < \frac{5\pi}{3}$, and the outflow condition ($\displaystyle
 \frac{\partial\bm{u}}{\partial r}=\bm{0}$ and $p=0$) was applied in the region $\displaystyle 0 \le \theta < \frac{\pi}{3}, \frac{5\pi}{3} < \theta < 2\pi$.

The computational domain was discretized to $N_r \times N_\theta$ grids, where $N_r$ is the division number in the radial direction and $N_\theta$ is the division number in the azimuthal direction. The grid was constructed such that the grid spaces were finer near the cylinder and in the downstream area (Fig. \ref{fig:Computational grid for the caculation of karman vortex street}). For computation, the Navier-Stokes equation was discretized by the finite volume method. The advection term was calculated using the flux splitting method\cite{liu98_numer_study_insec_fligh} with flux of the third order at the boundary of the control volume, and the dissipation term was calculated by the Crank-Nicolson method. The linear equations for the Poisson equation to obtain the pressure and the dissipation term were numerically solved by the BiCGSTAB method using the open software \textit{Lis} (\verb+https://www.ssisc.org/lis/+). For time integration, the Adams-Bashforth method was used. The periodic solution was obtained using the same algorithm as that applied for the traveling pulse of the FHN eqs in Sec.\ref{sec:A traveling pulse in Fitz-Hugh Nagumo equation in a periodic domain}.

The computational parameters are $N_r=N_\theta=60$ and $R=10$. In this condition, the radial grid width ranges from $0.00840$ to $0.308$, and the azimuthal grid width from $0.0525$ to $0.157$. The Reynolds number was $Re=200$. The phase sensitivity functions were calculated with coarser mesh, $N_r=N_\theta=40$, and found no substantial difference between the results with these meshes.

To calculate the periodic solution representing the K\'arm\'an's vortex street, the period was discretized to 1,000 steps. The time origin $t=0 (\phi=0)$ was set at the minimum of the lift coefficient. Eight snapshots of the obtained periodic solution are shown in Fig. \ref{fig:Karman vortex}. The periodic solution gives the mean drag coefficient $\langle C_D \rangle =\langle 2F_x \rangle=1.346$ and the Strouhal number $St=fD/U=0.201$. These values are close to the values of previous studies\cite{%
henderson95_detail_drag_curve_near_onset_vortex_shedd,%
williamson96_vortex_dynam_cylin_wake%
}.

\begin{figure}[h]
\centering
\includegraphics[width=0.6\textwidth]{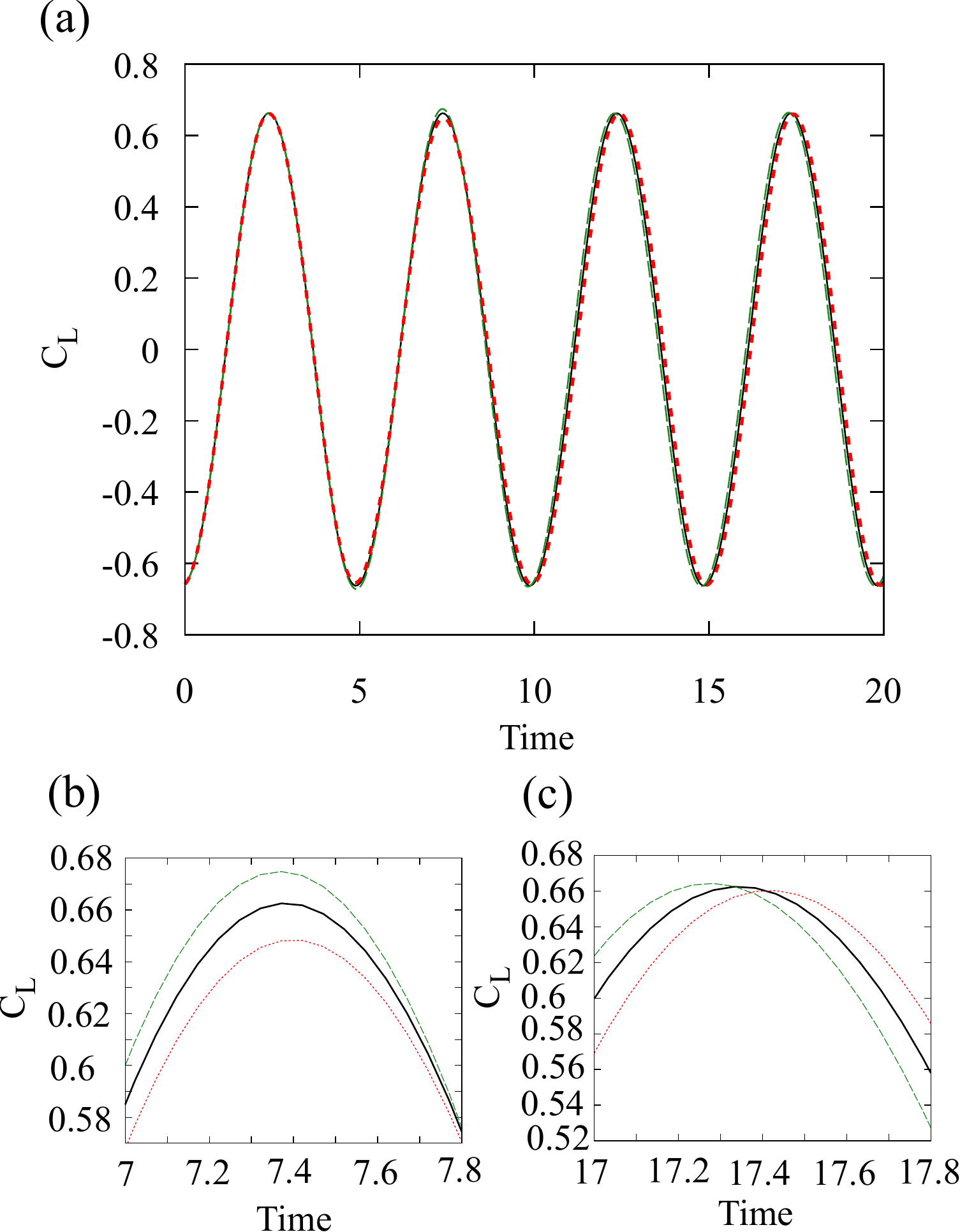}
\caption{
(a) Times series of the lift coefficient $C_L(t)$ of the perturbed and unperturbed cases. $C_L(t)$ of the unperturbed system is shown by the straight black line. $C_L(t)$ of the perturbed system are shown by the green dashed line ($\epsilon_0=0.5$) and red dotted line ($\epsilon_0=-0.5$), respectively. (b) Magnified graph of (a) in $7 \le t \le 7.8$. The peak heights of $C_L(t)$ are different and the peak positions (phases) are not significantly different. (c) Magnified graph of (a) in $17 \le t \le 17.8$. The peak heights become uniform due to the converging process to the limit cycle and the peak positions (phases) are shifted.
}
\label{fig:A demonstration of the phase shift}
\end{figure}

Fig. \ref{fig:A demonstration of the phase shift} shows the time series of $C_L$ for the periodic solution and those started from the perturbed states, to demonstrate the occurrence of phase shift. A perturbation was applied to the single horizontal velocity component of the periodic solution. The position $\bm{x}_{pert}$ was set $(-6.00,0)$, a point in the upstream of the cylinder, which is indicated by the red point in Fig. \ref{fig:Computational grid for the caculation of karman vortex street}. To perturb the velocity, the discretized velocity component was changed at $\bm{x}=\bm{x}_{pert}$ as $u \mapsto u \pm \epsilon_0$ at $t=0$, where $\epsilon_0=0.5$, which indicates that the velocity in the corresponding control volume (area) is changed. The time series of $C_L$ in the cases with the perturbations $\pm \epsilon_0$ and that in the case with no perturbation are shown in Fig. \ref{fig:A demonstration of the phase shift}, where the cases of $+\epsilon_0$ and $-\epsilon_0$ cause the advance and delay of the phase, respectively. The changes in the phase and amplitudes of $C_L$ become apparent in the time regime $t \gtrsim 4$; the perturbation does not cause significant changes in $C_L$ before it reaches the cylinder. Further, amplitude changes rather than phase changes are apparent during $4 \lesssim  t \lesssim 8$, which implies that a diffused perturbation slightly enhanced (reduced) the flow speed around the cylinder to cause an increase (decrease) in the lift force. After the perturbation was advected downstream, the periodic state of the K\'arm\'an's vortex street recovers whereas the phase shift remained. These behaviors will be discussed in Sec. \ref{sec:Phase sensitivity functions for karman's vortex street} in detail with the results of the phase reduction theory.

\subsubsection{Phase sensitivity functions for \karman's vortex street}
\label{sec:Phase sensitivity functions for karman's vortex street}
\begin{figure}[h]
\centering
\includegraphics[width=1.0\textwidth]{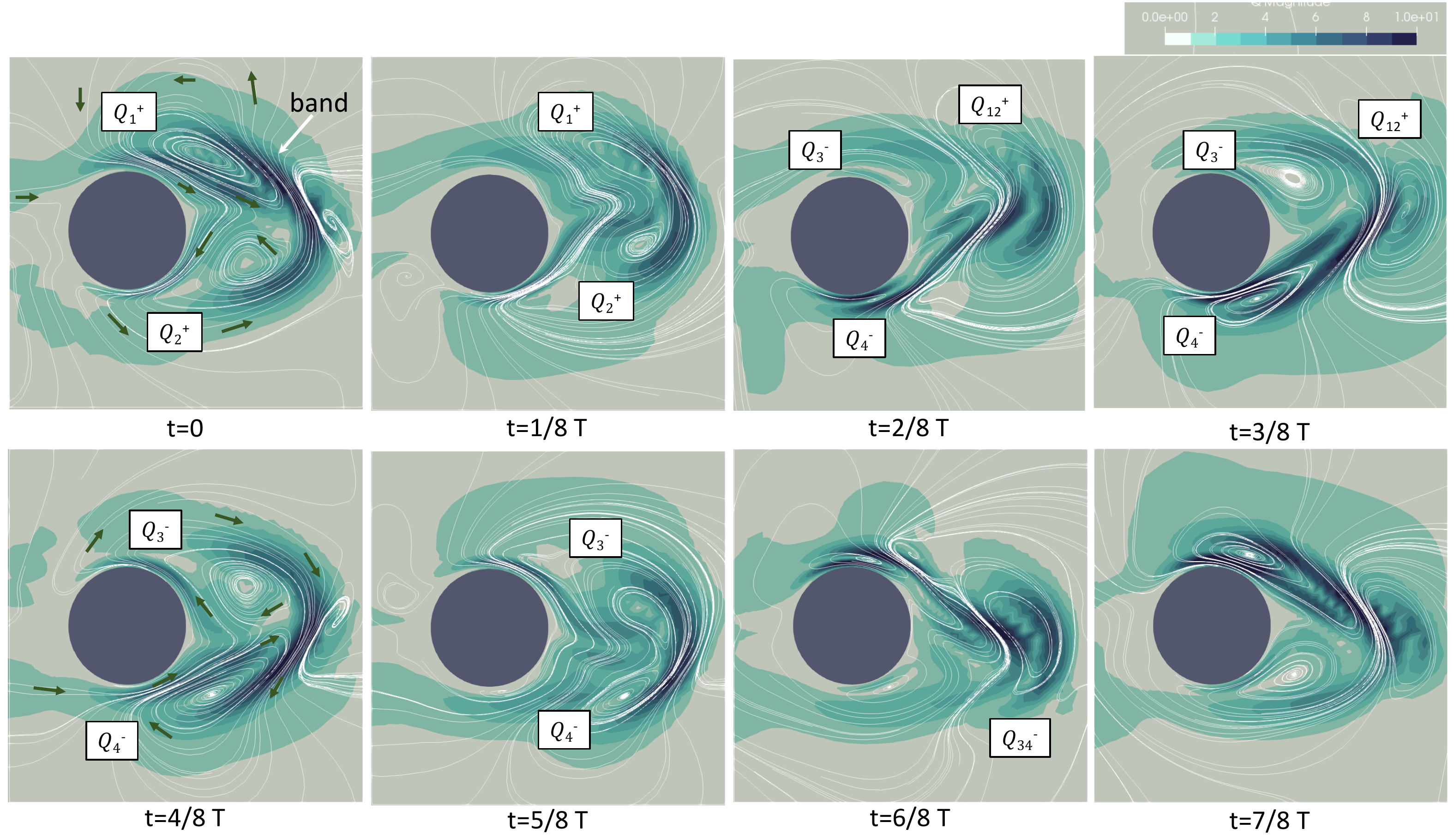}
\caption{
Phase sensitivity vector $\bm{Q}$ in the downstream region $[-1,3]\times[-2,2]$. Colors indicate $|\bm{Q}|$, and curves indicate the integration line of $\bm{Q}$. Green arrows indicate the direction of $\bm{Q}$, but the length does not imply anything. Some Q-eddies are labeled such as $Q_1^+$; the superscript indicates the direction of rotation.
}
\label{fig:Phase sensitivity vector}
\end{figure}

The proposed method enables us to produce the phase sensitivity function of Karman's vortex street or the LC of the Navier-Stokes equation. For the following discussion, the phase sensitivity vector for the fluid system, $\bm{Q}(\bm{x};\phi)$, is defined as
\begin{equation}
  \bm{Q}(\bm{x};\phi) = (Q_u, Q_v), \quad
  Q_u(\bm{x};\phi)
 = \left. \frac{\delta \Phi(\bm{u})}{\delta u} \right|_{\bm{u}=\bm{u}_p(\bm{x},\phi/\omega)}, \quad
  Q_v(\bm{x};\phi)
 = \left. \frac{\delta \Phi(\bm{u})}{\delta v} \right|_{\bm{u}=\bm{u}_p(\bm{x},\phi/\omega)},
\end{equation}
 where $\bm{u}_p(\bm{x},t)$ is the time periodic solution corresponding to the Karman's vortex street. The components of the phase sensitivity vector are the phase sensitivity function defined in Ref. \cite{nakao14_phase_reduc_approac_to_synch}. The phase sensitivity vector causes the phase shift due to a point-wise perturbation to the velocity field. By definition of the functional derivatives, the following relationship holds:
\begin{equation}
\Phi(\bm{u}(\bm{x},t)+\bm{u}'\delta(\bm{x}-\bm{x}_0))
-\Phi(\bm{u}(\bm{x},t))
\simeq
\int \frac{\delta \Phi(\bm{u})}{\delta \bm{u}} \cdot \bm{u}'\delta (\bm{x}-\bm{x}_0) d\bm{x}
= \bm{Q}(\bm{x}_0;\phi) \cdot \bm{u}',
\label{eq:meaning of Q}
\end{equation}
 when $|\bm{u}'|$ is small. The normalization of the phase sensitivity vector is given by:
\begin{eqnarray}
\omega &=& \int \bm{Q}(\bm{x};\phi) \cdot \frac{\partial \bm{u}}{\partial t}(\bm{x}) d\bm{x} \\
&\simeq& 
\sum_{i,j} \bm{Q}(\bm{x}_{i,j};\phi) \cdot \frac{\partial \bm{u}}{\partial t}(\bm{x}_{i,j}) \Delta S_{i,j},
\label{eq:normalization of Q(NS)}
\end{eqnarray}
where $\bm{x}_{i,j}$ is the representative position of the control volume(area) indicated by $(i,j)$ ($i$ and $j$ are the radial index and azimuthal index, respectively), and $\Delta S_{i,j}$ is the area of the control volume. To estimate the phase sensitivity vector $\bm{Q}(\bm{x};\phi)$, $(Z_{i,j}^u, Z_{i,j}^v)=(\partial \phi/\partial u_{i,j}, \partial \phi/\partial v_{i,j})$ is defined, where $(u_{i,j}, v_{i,j})$ is the discretized velocity component at the position indicated by $(i,j)$. Then, the following formula which is similar to Eq. (\ref{eq:Def of Q}) is obtained:
\begin{equation}
  \bm{Q}(\bm{x}_{i,j};\phi) =\frac{\omega}{\Delta S_{i,j}}(Z_{i,j}^{u}, Z_{i,j}^{v}).
\label{eq:Def of Q(NS)}
\end{equation}

\begin{figure}[h]
\centering
\includegraphics[width=1.0\textwidth]{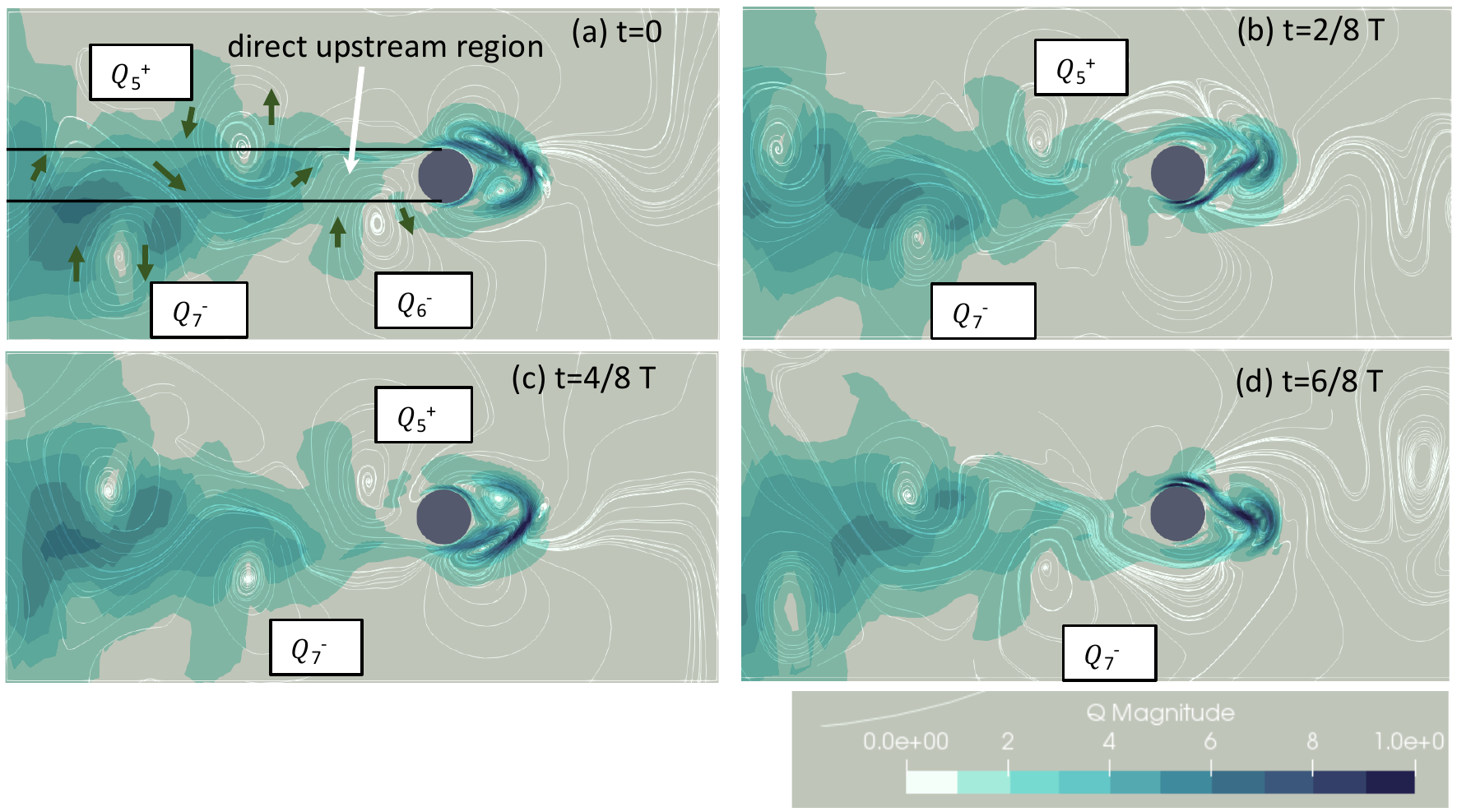}
\caption{
Similar to Fig. \ref{fig:Phase sensitivity vector}, but includes the upstream region; $[-8,5]\times[-3,3]$. The direct upward region and the region $|y|<1/2$ and $x<0$, are indicated. 
}
\label{fig:Phase sensitivity vector(wide).}
\end{figure}

In Figs. \ref{fig:Phase sensitivity vector} and \ref{fig:Phase sensitivity vector(wide).}, $\bm{Q}(\bm{x};\phi)$ is shown downstream of the cylinder and the wider region including the upstream region, respectively.

Downstream of the cylinder, a large value of $|\bm{Q}|$ is observed in the area whose size is approximately twice the diameter of the cylinder (green area in Fig. \ref{fig:Phase sensitivity vector}). One or two band(s) with particularly large $|\bm{Q}|$ ($|\bm{Q}|>6$) are observed in the downstream region of the cylinder, which change its shape with time (or phase).

The integral curves of $\bm{Q}$ in the large-$|\bm{Q}|$ region constructs several closed curves or spirals, which are referred to as ``Q-eddy'' henceforth. The direction of the large-$|\bm{Q}|$ band shares the edge of Q-eddy. During a single period, the Q-eddy is generated near the cylinder, and is transferred to the downstream and disappears. At $t=0$, two Q-eddies $Q_1^+$ and $Q_2^+$, both of which are counterclockwise, exist near the cylinder; this direction can be defined as positive and vice versa. These Q-eddies are transferred to the downstream to merge with the single Q-eddy $Q_{12}^+$ at $t=2/8T$. At the same time, new two Q-eddies $Q_3^-$ and $Q_4^-$ were detached from the cylinder. They are also transferred to the downstream to form $Q_{34}^-$ ($t=6/8T$).

When the large-$|\bm{Q}|$ band is compared with the flow speed distribution in Fig. \ref{fig:Karman vortex}, the band corresponds to the region where the edge of the low flow speed region at the back of the cylinder. Further, the direction corresponds to the flow vector of the eddy which is attached to the cylinder and will be detached (e.g. lower eddy in Fig.\ref{fig:Karman vortex}, $t=0$). This observation matches our expectation that such a perturbation supporting the vortex generation results in earlier separation leading to the phase advance. However, it must be noted that the structure depends significantly on both time and space, and the application of the phase control in this region requires regulating  the perturbation distribution to the velocity field.

In the upstream of the cylinder, the region with large $|\bm{Q}|$ spreads wider, although the peak values are not significantly high as those in the downstream (Fig. \ref{fig:Phase sensitivity vector(wide).}). The region has a triangular shape. A significant feature is that the $x-$component of $\bm{Q}$ in the direct upstream region of cylinder, which may be characterized by $|y| < D/2$ and $x<0$, is always positive. The phase shift due to the perturbation in this region is demonstrated in Sec.\ref{sec:Methods(karman)} (Fig.\ref{fig:A demonstration of the phase shift}).

These characteristics of the $\bm{Q}$ field can be explained as follows: The positive perturbation (e.g. $\bm{u}'=(\epsilon,0)$, where $\epsilon$ is a small positive number) in the upstream dissipates to spread during transfer to the downstream, which enhances the local speed near the cylinder. Because the period of the LC for the Karman's vortex street is related to the uniform flow by a constant Strouhal number, $St=fD/U$, the frequency $f$ is slightly enhanced due to a slight enhancing of the local speed. After the perturbation is transferred downstream, the state converges to the LC, but the existence of a time interval with large frequency shifts the phase in advance. For this scenario, the perturbation position should be located at a certain distance from the cylinder; such distance is needed to spread the perturbation so that it can be regarded as an enhancement of the local speed around the cylinder. These characteristics of the $\bm{Q}$ field suggest that a simple control strategy is possible by using the direct upstream region of the cylinder than other regions.

The Q-eddies in the upstream are located at a constant interval on both the sides of the direct upstream region (e.g. $Q_{6}^-$ and $Q_{7}^-$ in Fig. \ref{fig:Phase sensitivity vector(wide).}). They travel downstream as the time (phase) increases (for example, $Q_{5}^+$ in Fig. \ref{fig:Phase sensitivity vector(wide).}(a)-(c)). The spiral structure implies that the sign of the phase shift due to the constant perturbations in this area changes with time. The perturbation added to this region (upper or lower regions of the direct upstream region) is transferred to the side of the cylinder. Therefore, it works to change the separation timing rather than to enhance the flow speed around the cylinder. Thus, the Q-eddy structure should be spatially periodic with the period length roughly estimated by $T U \simeq 5$.

It is important to note that such property of the Q-eddy structure in the upstream is similar to the result in Sec. \ref{sec:A traveling pulse in Fitz-Hugh Nagumo equation in a periodic domain}, where $Q_u$ and $Q_v$ of the traveling pulse oscillated in the ``upstream'' of the pulse with oscillatory tail, though the ``upstream'' for the traveling pulse must be interpreted as the region where the pulse travels.

\begin{figure}[h]
\centering
\includegraphics[width=0.5\textwidth]{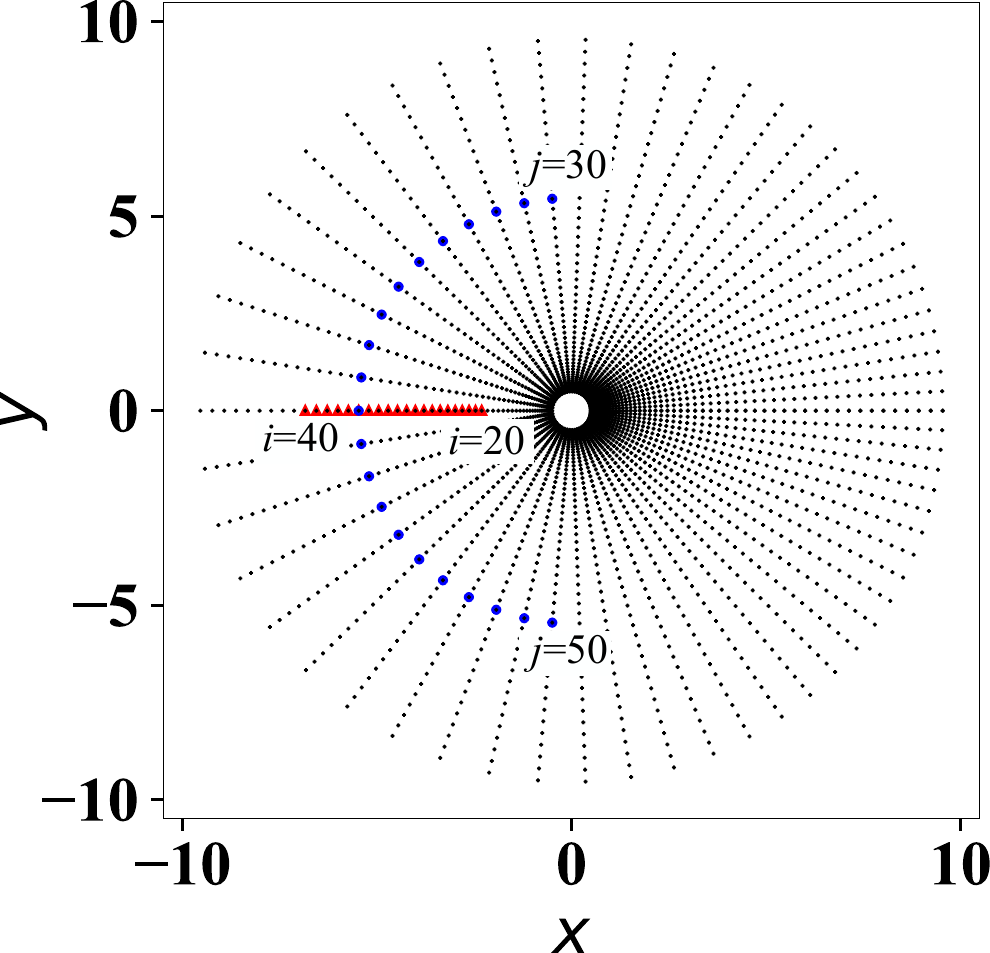}
\caption{
Test positions to compare the proposed method with the direct method: $\{(i,j)\mid 20 \le i \le 40, j=30\}$ (red triangles) and $\{(i,j)\mid i=45, 30 \le j \le 50\}$ (blue points), where $i$ and $j$ indicate the radial and azimuthal indices, respectively.
}
\label{fig:Test positions to check the phase sensitivity function}
\end{figure}

The phase sensitivity function calculated by the proposed method was compared with the result obtained by the direct method. Several points in the upstream shown in Fig. \ref{fig:Test positions to check the phase sensitivity function} were selected, which were used for the comparison.

The procedure of the direct method is as follows: the field at $t=0 (\phi=0)$ was chosen, and the discretized velocity component at each selected point, $(u,v)$, was perturbed by either $u \mapsto u+\epsilon_0$ or $v \mapsto v+\epsilon_0$. The calculation of the time integration started from the perturbed field to obtain the time series of $C_L$. Let us define $f_u(t)$ and $f_p(t)$ as the time series of $C_L$ for the unperturbed case and that for the perturbed case, respectively. Then, a $L^2-$norm of the difference between these two data, $\displaystyle
  \int_{-T/2}^{T/2}\left( f_u(t+t_0) - f_p(t)\right)^2 dt
$, was used to find the minimizer $t_0 (\in [-T/2,T/2))$ such that the $L^2-$norm was minimized. Then, the phase difference between these two data, $\Delta \phi$, is estimated as $\displaystyle
  \Delta \phi = \left(\frac{2\pi}{T}\right) t_0
$. For the comparison,  $\tilde{Z}_{i,j}^{u}(\phi/\omega)$ and $\tilde{Z}_{i,j}^{v}(\phi/\omega)$ are used, which are defined as:
\begin{equation}
Z_{i,j}^u(\phi) = \left(\frac{2\pi}{T}\right) \tilde{Z}_{i,j}\left(\frac{\phi}{\omega}\right).
\end{equation}
Thus, the phase difference by the direct method, $\Delta \phi$, is related to $Z_{i,j}^u(\phi)$ by the formula $\tilde{Z}_{i,j}^u(\phi/\omega)= t_0/\epsilon_0$, and similar formula for $Z_{i,j}^v(\phi/\omega)$. The shifted data was generated by Fourier transformation and the phase shift of the Fourier coefficients, and the minimizer was obtained by the downhill simplex method. The one period length of data has been during $[9T,10T]$.

\begin{figure}[h]
\centering
\includegraphics[width=1.0\textwidth]{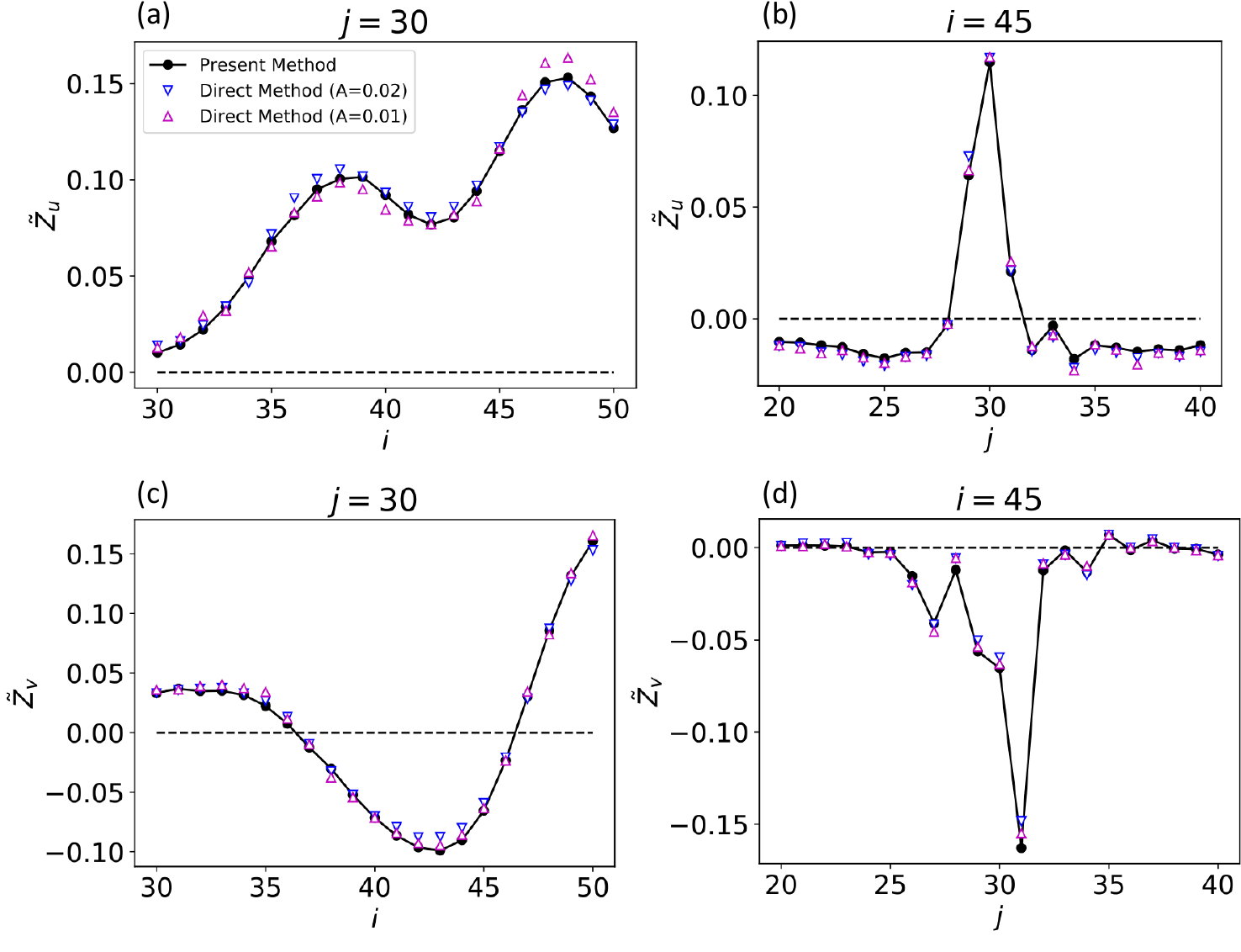}
\caption{Comparison between the present method and direct method. Test positions are indicated by red triangles and blue points in Fig. \ref{fig:Test positions to check the phase sensitivity function}. (a) $\tilde{Z}_u(\phi)$ along the red triangles, (b) $\tilde{Z}_u(\phi)$ along the blue points, (c) $\tilde{Z}_v(\phi)$ along the red triangles, and (d) $\tilde{Z}_v(\phi)$ along the blue points.
}
\label{fig:phase sensitivity function in the upstream}
\end{figure}

Figs. \ref{fig:phase sensitivity function in the upstream}(a)-(d) show $\tilde{Z}_{i,j}^u$ and $\tilde{Z}_{i,j}^v$ obtained by the proposed method and direct method, and the direct method with $\epsilon_0=0.01$ and $0.02$. Figs. \ref{fig:phase sensitivity function in the upstream}(a) and (c) show the values along the line segment indicated by red triangles in Fig. \ref{fig:Test positions to check the phase sensitivity function}, whereas Figs. \ref{fig:phase sensitivity function in the upstream}(b) and (d) show the values along the line segment indicated by blue points in Fig. \ref{fig:Test positions to check the phase sensitivity function}. In all cases, the agreement of the values between the present method and the direct method, as well as the convergence of the direct method with different perturbation amplitudes, is reasonable. It is remarkable that the phase shift represented by the ratio with the period, $t_0/ T$, is given $\epsilon \tilde{Z}_{i,j}^u/T$. Considering the magnitudes of $\tilde{Z}_{i,j}^u$ and $\tilde{Z}_{i,j}^v(\phi)$ are at most $0.15$, which indicates that $t_0/T \approx 0.003$ ($0.3\%$) or less if $\epsilon_0=0.01$ and $T \approx 5$. The value of the phase shift can be increased if the wider region is perturbed for a longer time interval according to the information of $\bm{Q}(\bm{x},\phi)$.

\section{Summary}\label{sec:Summary}
In this paper, a method to calculate the phase sensitivity function was developed, which is a fundamental function of the phase reduction theory. This method does not use the explicit form of the linearized matrix around the limit cycle (the Jacobian), which can be applied for the incompressible fluid system, where the Jacobian is determined by the Poisson equation. This method does not need the long-time integration until convergence like the direct method and the adjoint method, which reduces the computation time as well as the memory. Further, two applications were demonstrated: traveling pulse of the FitzHugh Nagumo equation in a periodic domain to validate this method, and the K\'arm\'an's vortex street, to demonstrate the application to incompressible fluid systems.

The phase sensitivity function for the K\'arm\'an's vortex street indicated how the phase shifts due to the external perturbation. Our analysis suggested that the phase shift property can be easily designed in the direct upstream region of the cylinder, where positive perturbations to the $x-$component of the velocity causes the phase advance, regardless of the phase in a wide area. This can be explained by a local speed-up due to the spread of the perturbative flow and constant Strouhal number for the K\'arm\'an's vortex street. Higher values are obtained in the downstream area of the cylinder; however, the effective region is narrow and phase-dependent, which suggests that the control of the phase requires detailed design of the perturbation distribution and direction. This result is fundamental in controlling the phase of the K\'arm\'an's vortex street.

The phase description is a powerful tool to analyze the phenomena with the limit cycle. For a large system, the numerical method to calculate the phase sensitivity function proposed here will be of great use, especially when the synchronization or the entrainment is considered. Further applications will be reported in future studies.

\begin{acknowledgements}
This work was partially supported by the Mazda foundation.
The author would like to thank Prof. Hiroya Nakao and Prof. Yoji Kawamura for discussions. 
\end{acknowledgements}

%

\end{document}